\documentclass[twocolumn]{aastex631}
\makeatletter
\renewcommand\@makefntext[1]{\leftskip=2em\hskip-2em\@makefnmark#1}
\makeatother
\usepackage{amsmath}

\shortauthors{M. De Furio et al.}
\shorttitle{A-type Binaries with CHARA}

\begin{document}
\title{The Small Separation A-Star Companion Population: Tentative Signatures of Enhanced Multiplicity with Primary Mass}

\correspondingauthor{Matthew De Furio}
\email{defurio@utexas.edu}

\author[0000-0003-1863-4960]{Matthew De Furio}
\affiliation{Department of Astronomy, The University of Texas at Austin, 2515 Speedway, Stop C1400, Austin, TX 78712, USA}
\affiliation{NSF Astronomy and Astrophysics Postdoctoral Fellow}

\author[0000-0002-3003-3183]{Tyler Gardner}
\affiliation{Astrophysics Group, Department of Physics and Astronomy, University of Exeter, Stocker Road, Exeter, EX4 4QL, UK}

\author[0000-0002-3380-3307]{John D. Monnier}
\affiliation{Department of Astronomy, University of Michigan, Ann Arbor, MI 48109, USA}

\author[0000-0003-1227-3084]{Michael R. Meyer}
\affiliation{Department of Astronomy, University of Michigan, Ann Arbor, MI 48109, USA}

\author[0000-0001-5253-1338]{Kaitlin M. Kratter}
\affiliation{Department of Astronomy and Steward Observatory, University of Arizona, Tucson, AZ}

\author[0000-0001-9745-5834]{Cyprien Lanthermann}
\affiliation{The CHARA Array of Georgia State University, Mount Wilson Observatory, Mount Wilson, CA 91203, USA}


\author[0000-0002-2208-6541]{Narsireddy Anugu}
\affiliation{The CHARA Array of Georgia State University, Mount Wilson Observatory, Mount Wilson, CA 91203, USA}

\author[0000-0001-6017-8773]{Stefan Kraus}
\affiliation{Astrophysics Group, Department of Physics and Astronomy, University of Exeter, Stocker Road, Exeter, EX4 4QL, UK}


\author[0000-0001-9745-5834]{Benjamin R. Setterholm}
\affiliation{Department of Astronomy, University of Michigan, Ann Arbor, MI 48109, USA}
\affiliation{Max-Planck-Institut für Astronomie, Heidelberg, DE}


\begin{abstract}
We present updated results from our near-infrared long-baseline interferometry (LBI) survey to constrain the multiplicity properties of intermediate-mass A-type stars within 80 pc. Previous adaptive optics surveys of A-type stars are incomplete at separations $<$ 20au. Therefore, a LBI survey allows us to explore separations previously unexplored. Our sample consists of 54 A-type primaries with estimated masses between 1.44-2.93 M$_{\odot}$ and ages 10-790 Myr, which we observed with the MIRC-X and MYSTIC instruments at the CHARA Array. We use the open source software CANDID to detect two new companions, seven in total, and we performed a Bayesian demographic analysis to characterize the companion population. We find the separation distribution consistent with being flat, and we estimate a power-law fit to the mass ratio distribution with index -0.13$^{+0.92}_{-0.95}$ and a companion frequency of 0.25$^{+0.17}_{-0.11}$ over mass ratios 0.1-1.0 and projected separations 0.01-27.54au. We find a posterior probability of 0.53 and 0.04 that our results are consistent with extrapolations based on previous models of the solar-type and B-type companion population, respectively. Our results suggest that the close companion population to A-type stars is comparable to that of solar-types and that close companions to B-type stars are potentially more frequent which may be indicative of increased disk fragmentation for stars $\gtrsim$ 3M$_{\odot}$.

\end{abstract}

\section{Introduction} \label{sec:intro}
The majority of Sun-like stars in the Galaxy have a stellar companion \citep{Raghavan2010, Moe2017}
The same may be true of low-mass M-type stars, the most common type of star, when also considering brown dwarf companions \citep{Winters2019AJ....157..216W, Susemiehl2022A&A...657A..48S}. The ubiquity of companions over a broad range of stars is a result of the star formation process. Companions form at early times ($\lesssim$ 1 Myr) and exist in stable configurations across orders of magnitude in primary mass (O-stars to brown dwarfs), orbital separation ($<$ 1 - 1000s au), and mass ratios. 

Companions to stars are believed to be predominantly produced through two common channels, disk fragmentation and turbulent fragmentation, resulting in various mass ratios and orbital separations as a function of host star mass. Disk fragmentation results in companions at separations less than the size of the disk, $\sim$ 100 au \citep{Adams1989, Bonnell1994}, while turbulent fragmentation of molecular cloud cores generates gravitationally bound companions out to 1000s of au \citep{Goodwin2004, Offner2010}. Subsequent processes can alter their separations through migration and cluster dynamics, and mass ratios through preferential accretion onto the companion \citep{Bate2002MNRAS.336..705B, Bate2003MNRAS.339..577B, YoungClarke2015MNRAS.452.3085Y, Offner2023ASPC..534..275O}. 
These processes can significantly affect circumstellar disk evolution, planet formation, and allowable planetary architectures \citep{Kraus2016AJ....152....8K, Moe2021MNRAS.507.3593M}. The outcomes of multiple formation place substantial constraints on star and planet formation theory, and a thorough understanding of multiplicity is necessary to develop a comprehensive theory of star formation.

Previous multiplicity surveys in the Galactic field demonstrate that the companion frequencies and distributions of mass ratio (M$_{\rm 2}/M_{\rm 1}$) and orbital separation depend on primary mass \citep{Offner2023ASPC..534..275O}. They find that the companion separation distribution can be modeled as a log-normal distribution with a peak at $\sim$ 6 au for brown dwarfs \citep{Reid2006}, $\sim$ 20 au for M-type primaries \citep{Janson2012, Winters2019AJ....157..216W}, and $\sim$ 50 au for solar-type primaries \citep{Raghavan2010}. Higher mass O and B stars appear to have separation distributions flat in log-space \citep{Moe2017}, with high frequencies of companions across all separations \citep{Offner2023ASPC..534..275O}. A-type primaries ($\sim$ 1.5-2.5 M$_{\odot}$) appear to have a peak around 390 AU \citep{DeRosa2014}. However, this is based on a multiplicity survey that suffers from incompleteness for separations $<$ 20 AU, leaving open the possibility for a significant population of close separation companions to A-type primaries \citep{Moe2017}.

The radial velocity (RV) survey of \citet{Carquillat2007} of Am stars, chemically peculiar A-stars with small rotational velocities, found a different peak of very close companions with a mean period of 5 days ($\sim$ 0.1 AU). Due to their small rotational velocity, these chemically peculiar A-stars are biased towards close companions that may reduce the rotational velocity of the primary, and therefore not representative of all A-type stars. A similar peak has been identified for O-type primaries \citep{Sana2012}, while a recent long-baseline interferometry survey finds a preference for companions near $\sim$ 10 au although potentially biased against closer companions \citep{Lanthermann2023A&A...672A...6L}. RV surveys of chemically typical A-type stars are difficult as they are usually fast rotators and have broad absorption lines, as opposed to the overwhelmingly slow rotational velocities of chemically peculiar A-type stars. Therefore, this type of survey has not been performed to date. In \citet{DeFurio2022_Atype} (hereafter DF22), we observed 27 typical A-type stars with LBI at the CHARA Array \citep{tenBrummelaar2016SPIE.9907E..03T} and found a handful of close companions. This sample is quite small and suffers from large error bars. Therefore, the close companion population around typical A-type stars is still underexplored. 
Expanding upon this sample is necessary to probe the impact of primary mass on the formation and evolution of close binary systems and filling in the gap in our understanding between solar-type and OB-type multiplicity. 

In this paper, we present the results of our expanded survey including new observations and an updated Bayesian demographic analysis. In Section 2, we describe the newly observed data and the methods to identify companions. In Section 3, we present the companion detections, describe our detection limits, and characterize the close companion population of A-type primary stars with a Bayesian analysis. In Section 4, we compare our results to various multiplicity surveys and discuss the implications. In Section 5, we summarize our conclusions.

\section{Methods} \label{sec:methods}

\subsection{Observations} \label{subsec:observations}

As a continuation of our A-star survey, we observed 30 additional chemically typical A-type stars over the course of four nights (UT May 6-8 and July 30, 2022) at the Center for High Angular Resolution Astronomy (CHARA) Array \citep{tenBrummelaar2005ApJ...628..453T}, an optical and near-infrared interferometric array consisting of six 1-meter telescopes with baselines up to 331 meters. All sources were observed over the H-band with the grism mode (R $\sim$ 190) on the Michigan Infra-Red Combiner-eXeter (MIRC-X) instrument \citep{Monnier2006SPIE.6268E..1PM, Anugu2020} and some over the K-band with the prism mode (R $\sim$ 50) on the Michigan Young Star Imager at CHARA (MYSTIC) instrument \citep{Monnier2018SPIE10701E..22M, Setterholm2023JATIS...9b5006S} once it became available to the public, see Table \ref{tab:table1}. The higher spectral resolution of MIRC-X allows us to identify companions at wider separations, out to roughly 0.3'' \citep{Anugu2020} while MYSTIC is used to confirm marginal detections and obtain color information on each detected companion. We also employed the etalon to properly calibrate the wavelength, see \citet{Gardner2021}. A typical observing sequence follows 10 min source integration, the standard shutters sequence \citep{Anugu2020}, and repeated 10 min source integration. For some bright primaries, we instead do 5+5 min source integrations for comparable S/N to all other targets. We opted to observe no calibrator stars as the majority of targets in our previous run were identified as single stars. Therefore, we use science targets well fit by a single star model to serve as calibrators to stars observed close in time, see Table \ref{table2}.

\begin{deluxetable*}{ccccccccc}
\tablenum{1}
\tablecolumns{9} 
\tablecaption{Table of sources observed with MIRC-X, and MYSTIC where indicated, in our full sample of 54 stars with Modified Julian Date (MJD) of observation.  Listed spectral types, ages, and masses for each star were taken from \citet{DeRosa2014} who describe their method of estimating age and mass in their Appendix.  Distances and their errors (16\% and 84\% confidence level) were extracted from the Gaia DR3 archive \citep{TheGaiaMission2016,Gaia_Multiplicity_2022arXiv220605595G, Babusiaux2022arXiv220605989B}, except where noted. Listed is the acceleration SNR from \citet{Kervella2022} that indicates deviations from the proper motion over the time baseline of Hipparcos and Gaia eDR3. Values greater than 3 are indicative of a likely companion over a broad range of potential separations and mass ratios. The non-single star catalog codes from Gaia DR3 are displayed, whose descriptions can be found in \citet{Gaia2022yCat.1357....0G}. }
\tablehead{\colhead{HD Number} &\colhead{HIP Number} &\colhead{SpType} &\colhead{Distance} &\colhead{Age}  &\colhead{Mass}   &\colhead{Date Observed} &\colhead{SNR acceleration} &\colhead{Gaia NSS} \\ \colhead{} &\colhead{} &\colhead{} &\colhead{(pc)} &\colhead{(Myr)}  &\colhead{(M$_{\odot}$)}   &\colhead{MJD}&\colhead{(Kervella+22)}&\colhead{Code}}
\startdata
1404*$^{\#}$	  &	1473	 &	A2V	 &	43.00$^{+0.34}_{-0.33}$	&	200	&	2.26	 &	59790.43	 &	1.16	 &	       None	\\
4058*$^{\#}$	  &	3414	 &	A5V	 &	54.77$^{+3.39}_{-0.65}$	&	250	&	1.9	 &	59790.45	 &	1.47	 &	     tbosb2	\\
5448	  &	4436	 &	A5V	 &	42.37$^{+0.11}_{-0.19}$	&	450	&	2.39	 &	59203.08	 &	1.82	 &	       None	\\
11636$^{\dagger}$	  &	8903	 &	A5V	 &	18.27$^{+0.25}_{-0.25}$	&	630	&	2.01	 &	59204.08	 &	      N/A	 &	       None	\\
14055*$^{\#}$	  &	10670	 &	A1Vnn	 &	35.74$^{+0.60}_{-0.58}$	&	100	&	2.57	 &	59790.48	 &	1.19	 &	       None	\\
15550	  &	11678	 &	A9V	 &	71.77$^{+1.84}_{-0.56}$	&	790	&	1.84	 &	59204.14	 &	1.28	 &	       None	\\
20677	  &	15648	 &	A3V	 &	48.11$^{+0.30}_{-0.25}$	&	250	&	2.11	 &	59203.18	 &	1.20	 &	       None	\\
21912	  &	16591	 &	A3V	 &	56.29$^{+0.16}_{-0.15}$	&	40	&	1.77	 &	59203.16	 &	0.77	 &	     tbosb1	\\
24809	  &	18547	 &	A8V	 &	63.76$^{+0.12}_{-0.11}$	&	100	&	1.7	 &	59203.22	 &	4.34	 &	       None	\\
28910	  &	21273	 &	A8V	 &	46.97$^{+0.38}_{-0.37}$	&	630	&	2.21	 &	59204.22	 &	1.00	 &	    tbootsc	\\
29388$^{\dagger}$	  &	21589	 &	A6V	 &	47.1$^{+1.2}_{-1.2}$	&	630	&	2.17	 &	59204.19	 &	4.99	 &	       None	\\
31647*	  &	23179	 &	A1V	 &	49.91$^{+0.29}_{-0.29}$	&	30	&	2.39	 &	59508.56	 &	0.85	 &	       None	\\
32301	  &	23497	 &	A7V	 &	57.55$^{+4.80}_{-1.87}$	&	630	&	2.22	 &	59204.25	 &	0.59	 &	       None	\\
46089$^{\dagger}$	  &	31119	 &	A3V	 &	63.7$^{+1.5}_{-1.5}$	&	560	&	2.2	 &	59203.29	 &	1.19	 &	       None	\\
48097$^{\dagger}$	  &	32104	 &	A2V	 &	43.6$^{+1.3}_{-1.3}$	&	30	&	1.94	 &	59203.32	 &	0.58	 &	       None	\\
56537$^{\dagger}$	  &	35350	 &	A3V	 &	30.9$^{+0.2}_{-0.2}$	&	320	&	2.39	 &	59204.28	 &	0.90	 &	       None	\\
59037	  &	36393	 &	A4V	 &	55.91$^{+3.84}_{-1.44}$	&	500	&	2.16	 &	59203.35	 &	1.44	 &	       None	\\
66664	  &	39567	 &	A1V	 &	65.89$^{+0.70}_{-0.58}$	&	320	&	2.42	 &	59204.33	 &	0.74	 &	       None	\\
74198$^{\dagger}$	  &	42806	 &	A1IV	 &	55.6$^{+0.6}_{-0.6}$	&	320	&	2.49	 &	59204.39	 &	1.76	 &	       None	\\
74873	  &	43121	 &	 A1V	 &	54.82$^{+0.15}_{-0.14}$	&	50	&	1.88	 &	59204.36	 &	2.40	 &	       None	\\
77660	  &	44574	 &	A8V	 &	78.28$^{+0.20}_{-0.19}$	&	710	&	1.81	 &	59203.43	 &	2.26	 &	       None	\\
84107	  &	47701	 &	A2IV	 &	51.24$^{+2.38}_{-0.98}$	&	10	&	1.44	 &	59203.45	 &	1.16	 &	       None	\\
92941$^{\dagger}$	  &	52513	 &	 A5V	 &	66.9$^{+1.4}_{-1.4}$	&	450	&	1.84	 &	59204.48	 &	1.18	 &	       None	\\
97244	  &	54688	 &	A5V	 &	62.19$^{+0.18}_{-0.15}$	&	60	&	1.72	 &	59204.45	 &	2.11	 &	       None	\\
99787	  &	56034	 &	A2V	 &	69.64$^{+0.68}_{-0.89}$	&	280	&	2.32	 &	59203.48	 &	1.43	 &	       None	\\
106591$^{\#}$	  &	59774	 &	A3V	 &	24.86$^{+0.58}_{-2.58}$	&	320	&	2.31	 &	59706.22	 &	0.76	 &	    tbootsc	\\
106661	  &	59819	 &	A3V	 &	66.01$^{+0.09}_{-0.21}$	&	400	&	2.29	 &	59204.51	 &	1.44	 &	       None	\\
112734	  &	63320	 &	A5	 &	73.73$^{+0.44}_{-0.30}$	&	40	&	1.69	 &	59203.53	 &	0.64	 &	       None	\\
115271$^{\dagger}$	  &	64692	 &	A7V	 &	74.1$^{+2.4}_{-2.4}$	&	560	&	2.1	 &	59203.59	 &	65.69	 &	       None	\\
118232$^{\#}$	  &	66234	 &	A5V	 &	56.83$^{+0.36}_{-0.35}$	&	500	&	2.38	 &	59706.27	 &	1.31	 &	       None	\\
120047	  &	67194	 &	A5V	 &	52.84$^{+0.79}_{-0.39}$	&	500	&	1.78	 &	59204.6	 &	0.47	 &	       None	\\
121164	  &	67782	 &	A7V	 &	73.60$^{+0.58}_{-0.44}$	&	500	&	1.97	 &	59204.57	 &	1.56	 &	       None	\\
124675$^{\#}$	  &	69483	 &	A8IV	 &	49.62$^{+0.27}_{-0.27}$	&	500	&	2.38	 &	59706.31	 &	1.32	 &	       None	\\
125161$^{\#}$	  &	69713	 &	A7V	 &	29.86$^{+0.97}_{-0.35}$	&	50	&	1.81	 &	59706.33	 &	0.90	 &	       None	\\
130109$^{\#}$	  &	72220	 &	A0V	 &	43.46$^{+0.45}_{-0.42}$	&	320	&	2.71	 &	59705.32	 &	3.06	 &	       None	\\
\enddata
\tablenotetext{*}{Observed with both MIRC-X and MYSTIC at CHARA Array.}
\tablenotetext{\#}{ Newly observed in this work.}
\tablenotetext{\dagger}{Distance taken from \citet{DeRosa2014} using the Hipparcos catalog \citep{Hipparcos1997ESASP1200.....E} due to unreliable Gaia measurements for bright stars.}
\label{tab:table1}
\end{deluxetable*}

\begin{deluxetable*}{ccccccccc}
\tablenum{1}
\tablecolumns{9} 
\tablehead{\colhead{HD Number} &\colhead{HIP Number} &\colhead{SpType} &\colhead{Distance} &\colhead{Age}  &\colhead{Mass}   &\colhead{Date Observed} &\colhead{SNR acceleration} &\colhead{Gaia NSS} \\ \colhead{} &\colhead{} &\colhead{} &\colhead{(pc)} &\colhead{(Myr)}  &\colhead{(M$_{\odot}$)}   &\colhead{MJD}&\colhead{(Kervella+22)}&\colhead{Code}}
\startdata
141378$^{\#}$	&	77464	 &	A5IV	&	 53.77$^{+0.20}_{-0.19}$	&	250	&	2	&	59705.35	&	0.76	&	       None	\\
147547$^{\#}$	&	80170	 &	A9III	&	 61.84$^{+0.51}_{-0.56}$	&	500	&	2.5	&	59707.37	&	1.77	&	       None	\\
152107$^{\#}$	&	82321	 &	A2Vspe...	&	 53.31$^{+0.27}_{-0.27}$	&	400	&	2.31	&	59706.37	&	6.66	&	       None	\\
154494$^{\#}$	&	83613	 &	A4IV	&	 42.22$^{+0.20}_{-0.02}$	&	280	&	2.02	&	59707.41	&	1.29	&	       None	\\
156729*$^{\#}$	&	84606	 &	A2V	&	 52.08$^{+0.07}_{-0.06}$	&	350	&	2.4	&	59790.18	&	26.46	&	       None	\\
158352$^{\#}$	&	85537	 &	A7V	&	 64.03$^{+0.05}_{-0.07}$	&	630	&	2.1	&	59705.42	&	2.52	&	       None	\\
165777$^{\#}$	&	88771	 &	A4IVs	&	 27.72$^{+4.18}_{-0.60}$	&	400	&	2.08	&	59705.45	&	1.36	&	       None	\\
173582$^{\#}$	&	   XXX	 &	A3	&	 48.32$^{+0.96}_{-0.92}$	&	400	&	2.13	&	59706.41	&	      N/A	&	       None	\\
173607$^{\#}$	&	   XXX	 &	A5	&	 49.17$^{+0.25}_{-0.25}$	&	350	&	1.99	&	59706.45	&	      N/A	&	       None	\\
173880$^{\#}$	&	92161	 &	A5III	&	 27.95$^{+0.04}_{-0.03}$	&	110	&	1.94	&	59705.48	&	138.42	&	       None	\\
174602*$^{\#}$	&	92405	 &	A3V	&	 74.20$^{+1.55}_{-1.10}$	&	500	&	2.32	&	59790.22	&	0.94	&	       None	\\
177724$^{\#}$	&	93747	 &	A0Vn	&	 26.16$^{+0.24}_{-0.24}$	&	130	&	2.93	&	59705.51	&	1.18	&	       None	\\
184006$^{\#}$	&	95853	 &	A5V	&	 37.54$^{+0.24}_{-0.24}$	&	450	&	2.34	&	59706.49	&	1.02	&	       None	\\
192640*$^{\#}$	&	99770	 &	A2V	&	 40.74$^{+0.15}_{-0.15}$	&	40	&	2.71	&	59790.26	&	2.01	&	       None	\\
199254$^{\#}$	&	103298	 &	A4V	&	 61.29$^{+1.31}_{-0.52}$	&	450	&	2.06	&	59705.53	&	56.40	&	       None	\\
204414*$^{\#}$	&	105966	 &	A1V	&	 60.25$^{+0.39}_{-0.32}$	&	200	&	2.19	&	59790.3	&	3.13	&	       None	\\
205835*$^{\#}$	&	106711	 &	A5V	&	 67.33$^{+4.37}_{-3.14}$	&	630	&	2.2	&	59790.33	&	3.24	&	    tbootsc	\\
210715*$^{\#}$	&	109521	 &	A5V	&	 55.61$^{+0.65}_{-0.80}$	&	500	&	2.04	&	59790.37	&	80.77	&	       None	\\
213558*$^{\#}$	&	111169	 &	A1V	&	 31.75$^{+0.20}_{-0.20}$	&	130	&	2.45	&	59790.39	&	0.69	&	       None	\\
\enddata
\tablenotetext{*}{Observed with both MIRC-X and MYSTIC at CHARA Array.}
\tablenotetext{\#}{ Newly observed in this work.}
\tablenotetext{\dagger}{Distance taken from \citet{DeRosa2014} using the Hipparcos catalog \citep{Hipparcos1997ESASP1200.....E} due to unreliable Gaia measurements for bright stars.}
\end{deluxetable*}

\subsection{Data Reduction and Analysis} \label{subsec:astar_datareduction}
We followed the same data reduction steps as DF22 and summarized here. We used the standard MIRC-X data pipeline (version 1.3.3) \citet{Anugu2020} to measure the visibilities, closure phases, and differential phases from each baseline pair in the raw interferometric data. We used the default reduction parameters, and set coherent integration frames (ncoh) to 10 and maximum integration time to 60s to allow for wide companion detection. For final calibration, we used a modified version of the MIRC-X and original MIRC pipeline \citep{Monnier2007Sci...317..342M, zhao2009ApJ...701..209Z, che2011ApJ...732...68C, monnier2012} to filter out bad quality data by applying various quality checks, producing more reliable sensitivity maps \citep[e.g.][]{Gardner2021}.

We also followed the same data analysis routine as DF22 and summarized here. We utilized the open source python code CANDID \citep{Gallenne2015} to identify companions and define sensitivity limits to all of the sources in our sample from their measured interferometric observables, see \citet{Jennison1958MNRAS.118..276J,Rogers1974ApJ...193..293R, Monnier2000plbs.conf..203M, Monnier2004ApJ...602L..57M} for a description of these interferometric parameters. With CANDID, we fit single and binary star models to the closure phases measured for each source in order to identify companions because this interferometric observable is most sensitive to the presence of companions. For sources with companions, we then ran a more thorough analysis on the closure phases and squared visibilities to properly model the astrophysical scene, results of which are displayed in Table \ref{table3}.  We assume a uniform disk diameter in our model of both the primary and secondary. Some of these sources are likely rapid rotators which would make them elongated if resolved and the assumption of a uniform disk too simplistic. Even in this case, the best fit $\chi_{\nu}^{2}$ values to the detected companions still demonstrate that the model well represents the data. Additionally, for all companion detections we divide by a correction factor of 1.00535 to the fitted separation to bring the data to an absolute wavelength scale \citep{Gardner2022AJ....164..184G} with the corrected values in Table \ref{table3}.

\begin{deluxetable}{cccc}
\tablenum{2}
\tablecolumns{4} 
\tablecaption{Table of calibrators.}
\tablehead{\colhead{Calibrator} &\colhead{UD diameter} &\colhead{Night}  &\colhead{UD Reference} \\ \colhead{Name} &\colhead{(mas)} & \colhead{UT} & \colhead{} }
\startdata
HD 99787 & 0.303 $\pm$ 0.024 & 20 Dec. 2020 & 1 \\
HD 44851 & 0.58 $\pm$ 0.014 & 20 Dec. 2020 & 2 \\
HD 21912 & 0.276 $\pm$ 0.007 & 20 Dec. 2020 & 2 \\
HD 120047 & 0.308 $\pm$ 0.008 & 21 Dec. 2020 & 2 \\
HD 32301 & 0.479 $\pm$ 0.033 & 21 Dec. 2020 & 2 \\
HD 74198 & 0.362 $\pm$ 0.024 & 21 Dec. 2020 & 2 \\
HD 19066* & 0.85 $\pm$ 0.06 & 21 Oct. 2021 & 2 \\
HD 141378 & 0.314 $\pm$ 0.009 & 6 May 2022 & 2\\
HD 158352 & 0.41 $\pm$ 0.01 & 6 May 2022 & 2\\
HD 173880 & 0.47 $\pm$ 0.05 & 6 May 2022 & 2\\
HD 124675 & 0.55 $\pm$ 0.06 & 7 May 2022 & 2\\
HD 152107 & 0.39 $\pm$ 0.03 & 7 May 2022 & 2\\
HD 184006 & 0.59 $\pm$ 0.04 & 7 May 2022 & 3\\
HD 161868 & 0.57 $\pm$ 0.04 & 8 May 2022 & 2\\
HD 174602* & 0.36 $\pm$ 0.01 & 30 July 2022 & 2\\
HD 204414* & 0.282 $\pm$ 0.009 & 30 July 2022 & 2\\
HD 14055* & 0.47 $\pm$ 0.033 & 30 July 2022 & 3
\enddata
\tablenotetext{*}{Observed with both MIRC-X and MYSTIC at CHARA Array.}
\tablenotetext{1}{\citet{Swihart2017AJ....153...16S}}
\tablenotetext{2}{\citet{Bourges2017yCat.2346....0B}}
\tablenotetext{3}{searchcal \citep{Chelli2016}}
\label{table2}
\end{deluxetable}

\begin{deluxetable*}{cccccccc}
\tablenum{3}
\tablecolumns{9} 
\tablecaption{Detected binaries with flux ratio, projected separations in milli-arcseconds (corrected from absolute wavelength scale on MIRC-X using the scale factor of 1.00535, see Sec. \ref{subsec:astar_datareduction}), position angles in degrees, uniform disk diameter of the primary and secondary in milli-arcseconds, and the reduced chi squared test statistic of a single star model and binary star model.  All detections achieved the maximum significance threshold on CANDID, 8$\sigma$.}
\tablehead{
\colhead{Target} & \colhead{Flux Ratio} & \colhead{Projected} & \colhead{PA (deg)} & \colhead{UD$_{1}$} & \colhead{UD$_{2}$} & \colhead{$\chi^{2}_{\nu, 1}$} & \colhead{$\chi^{2}_{\nu,2}$} \\
\colhead{Name} & \colhead{} & \colhead{Sep. (mas)} & \colhead{(E of N)} & \colhead{(mas)} & \colhead{(mas)} & &  }
\decimalcolnumbers
\startdata
HD 5448$^{a}$ &  0.01573$^{+0.00039}_{-0.00042}$ & 64.785 $\pm$ 0.007 & 144.324 $\pm$ 0.007 & 0.695 $\pm$ 0.002 & - & 1.75 & 1.22 \\
HD 11636$^{a}$ & 0.13546$^{+0.0011}_{-0.00089}$ &  63.293 $\pm$ 0.002 &  102.271 $\pm$ 0.002 & 1.0819 $\pm$ 0.0008 & 0.549 $\pm$ 0.007 & 159.5 & 3.18\\
HD 28910$^{a}$ & 0.80146$^{+0.00064}_{-0.00067}$ &  6.102 $\pm$ 0.0013 &  304.81 $\pm$ 0.012 & 0.377$^{+0.007}_{-0.006}$ & 0.341$^{+0.008}_{-0.007}$ & 1332 & 1.17\\
HD 29388$^{a}$ & 0.02488$^{+0.00024}_{-0.00023}$ &  10.126 $\pm$ 0.002 &  23.59 $\pm$ 0.02 & 0.553 $\pm$ 0.005 & - & 4.39 & 1.14\\
HD 48097$^{a}$ & 0.0284$^{+0.00119}_{-0.0011}$ &  28.145 $\pm$ 0.007  & 14.16 $\pm$ 0.02 & 0.288 $\pm$ 0.015 & - & 2.26 & 1.51 \\
HD 205835 & 0.3126$\pm$0.0006 &  16.631 $\pm$ 0.001 & 181.55 $\pm$ 0.006 & 0.5  & - & 158.13  & 1.75 \\
HD 210715 & 0.0342$\pm$0.0003 & 157.296 $\pm$ 0.006  & 119.07 $\pm$ 0.012 & 0.5  & - & 2.37 & 1.48 \\
\enddata
\tablenotetext{a}{Fitted values taken from \citet{DeFurio2022_Atype}.}
\label{table3}
\end{deluxetable*}

\begin{deluxetable}{ccccc}
\tablenum{4}
\tablecolumns{5}z
\tablecaption{ Physical separation in au, masses (\(\textup{M}_\odot\)), and mass ratios (q).  Companion masses are estimated using the MIST evolutionary models and the assumed primary mass and age from \citet{DeRosa2014}, see Table \ref{tab:table1}.}

\tablehead{\colhead{Target}& \colhead{Physical} & \colhead{M$_{prim}$} & \colhead{M$_{sec}$} & \colhead{q}\\
\colhead{Name} & \colhead{Sep. (au)} & \colhead{(M$_{\odot}$)} & \colhead{(M$_{\odot}$)} &  }
\startdata
HD 5448$^{a}$ & 2.745 $\pm$ 0.013 & 2.39 &  0.60 & 0.25 \\
HD 11636$^{a}$ &  1.156 $\pm$ 0.013 & 2.01 & 1.05 & 0.52\\
HD 28910$^{a}$ &  0.287 $\pm$ 0.002 & 2.21 & 2.12 & 0.96\\
HD 29388$^{a}$ &  0.477 $\pm$ 0.012 & 2.17 &  0.67 & 0.31\\
HD 48097$^{a}$ &  1.23 $\pm$ 0.04  & 1.94 & 0.40 & 0.21\\
HD 205835 &  1.120 $\pm$ 0.002  & 2.20 & 1.57 & 0.71\\
HD 210715 &  8.747 $\pm$ 0.004  & 2.04 & 0.70 & 0.34
\enddata
\tablenotetext{a}{Fitted values taken from \citet{DeFurio2022_Atype} corrected using the scale factor in \citet{Gardner2022AJ....164..184G}.}
\label{table4}
\end{deluxetable}

\section{Results} \label{sec:results}

\subsection{Detections} \label{subsec:astar_detections}
We detected two new companions out of the 30 additional A-type stars in our sample, HD 205835 and HD 210715.  Both detections had a significance of 8$\sigma$, the maximum permitted value in CANDID, indicative of strong detections. These companions were detected at projected separations of 16.631 and 157.296 mas (physical separation of 1.120 and 8.747 au) with flux ratios = 31.3 and 3.4\%, respectively, see Table \ref{table3}. We estimated the masses of both new companions using the measured flux ratio of the system and the MIST evolutionary models \citep{Paxton2011ApJS..192....3P, Paxton2013ApJS..208....4P, Paxton2015ApJS..220...15P,Paxton2018ApJS..234...34P, Choi2016ApJ...823..102C, Dotter2016ApJS..222....8D}, assuming the primary mass and age from \citet{DeRosa2014}.  The mass ratios (q) of these systems are 0.71 and 0.34 respectively, see Table \ref{table4}. Both of these targets were reported to have a proper motion anomaly using Hipparcos and Gaia catalogs that could be indicative of a companion \citep{Kervella2019A&A...623A..72K, Kervella2022}. Several other sources in our sample were reported as having a high proper motion anomaly but without detections with CHARA, see Table \ref{tab:table1}, either due to the suspected companion outside of our resolution range (i.e. a $\gtrsim$ 30 au) or too faint given our contrast limits (i.e. q $\lesssim$ 0.15). HD 11636 and HD 28910 are two binaries that we previously detected with MIRC-X in \citet{DeFurio2022_Atype} and were previously spectroscopically identified \citep{Abt1965, Pourbaix2000A&AS..145..215P}. HD 205835 was identified in Gaia Data Release 3 (DR3) as a non-single star with the label ``OrbitalTargetedSearch'' meaning this companion was previously known and used to test the astrometric orbit fitting code \citep{Gaia_Multiplicity_2022arXiv220605595G, Holl2023A&A...674A..10H}. Gaia DR3 reports a period of 596 days and an eccentricity of 0.51. Given the adopted masses and period of the binary system, we estimate from Kepler's third law a semi-major axis of 2.16 au and a minimum and maximum distance from the primary as 1.06 and 3.26 au assuming the eccentricity from Gaia. Our observed projected separation given the distance to the source is 1.12 au which is within the estimated bounds. Of the other sources in our full sample, only HD 106591 was identified in Gaia DR3 as being a non-single star. It has an astrometrically identified companion with a period of 53 days and eccentricity of 0.45. Given its period and the distance to this source, the companion should have a semi-major axis of $\sim$ 0.21-0.56 au (8 - 23 mas), within our resolution sensitivity, but depending on orientation of the orbit and time of observation could be unresolvable. 
Without a companion mass estimate, we are unable to determine whether such a source should be detectable in the interferometric data. All other sources within our sample had no companion detections in either our analysis or Gaia DR3.

For these two new detections, the squared visibilities were noisy, and it was difficult to estimate the diameters and resolved flux. Therefore, we only used the closure phases to identify companions and estimate fluxes. However, there was no impact of a resolved primary on the derived companion parameters, and thus our exclusive use of the closure phases for these two sources is appropriate and the companion parameters are reliable.

\subsection{Detection Limits}\label{subsec:detectionlimits}
As in DF22, we used CANDID to define the limit at which we can recover true companions to each source within our sample. A thorough description is given in DF22, and we summarize here. CANDID contains a function that injects the signal of a companion of various flux ratios and positions into the interferometric observables. Then, it derives the flux ratio at each separation over a specified percentile of position angles (we select 99\%) where the detection significance level relative to a single star model equals 5$\sigma$ \citep{Gallenne2015}. For targets with detected companions, CANDID removes the best fit signal from the data and repeats the same process to define limits on the residual closure phases. Detection limits in terms of contrast are shown in Table \ref{table5}. We then calculate the mass of an object with the parameters of the derived detection limit using the age and mass of the primary from \citet{DeRosa2014} and the MIST evolutionary models. We then convert this into mass ratio, see Fig. \ref{fig:astar_survey_detetionprob}.

Importantly, CANDID assumes that errors in the interferometric observables are uncorrelated. However, interferometric observables are known to have correlated errors. \citet{Kammerer2020A&A...644A.110K} estimated that the attainable contrast increases by a factor of 2 when estimating correlations as opposed to assuming uncorrelated errors. While assuming uncorrelated errors is a common approach to analyzing interferometric data \citep[e.g.][]{Gallenne2015, Lanthermann2023A&A...672A...6L}, it must be stated that this will only provide a conservative estimate to our limits, and does not achieve the absolute attainable contrast limit. No software is available for the CHARA array that estimates correlations within the interferometric observables, and so we proceed with our analysis acknowledging that detection limits are conservative estimates.

\begin{deluxetable}{c|cccccc}
\tablenum{5}
\tablecolumns{7}
\tablecaption{Detection limits (contrast in units of magnitudes for the H band) derived for each of the new targets in our sample using MIRC-X at 1.0, 3.0, 5.0, 10.0, 50.0, and 300.0 milli-arcseconds (mas) in angular separation. We define the detection limit as the highest flux ratio companion that CANDID can recover at a given radius that is equivalent to a 5$\sigma$ detection. \label{sensitivity}}
\tablewidth{0pt}
\tablehead{\colhead{} & \colhead{[mas]} \\
\colhead{Target Name} & \colhead{1.0} & \colhead{3.0}& \colhead{5.0}& \colhead{10.0}& \colhead{50.0}& \colhead{300.0}}
\startdata
HD 1404	&	4.41	&	4.41	&	5.19	&	5.28	&	5.57	&	5.05	\\
HD 4058	&	3.48	&	3.48	&	4.07	&	4.12	&	4.59	&	4.03	\\
HD 14055	&	4.05	&	4.31	&	4.93	&	5.17	&	5.48	&	4.96	\\
HD 106591	&	3.97	&	5.21	&	5.22	&	5.50	&	5.52	&	4.98	\\
HD 118232	&	2.45	&	3.77	&	3.93	&	4.16	&	4.34	&	3.63	\\
HD 124675	&	3.05	&	4.76	&	4.76	&	5.08	&	5.18	&	4.57	\\
HD 125161	&	3.01	&	3.44	&	3.44	&	3.98	&	4.53	&	4.01	\\
HD 130109	&	3.82	&	3.82	&	3.82	&	4.78	&	5.19	&	4.72	\\
HD 141378	&	3.71	&	3.75	&	3.75	&	3.76	&	4.42	&	3.89	\\
HD 147547	&	0.98	&	0.98	&	1.34	&	2.50	&	2.86	&	2.35	\\
HD 152107	&	3.63	&	4.04	&	4.14	&	4.36	&	4.55	&	3.91	\\
HD 154494	&	1.98	&	1.98	&	1.98	&	3.20	&	4.00	&	3.33	\\
HD 156729	&	1.01	&	1.01	&	1.01	&	2.12	&	2.53	&	2.00	\\
HD 158352	&	3.80	&	3.80	&	3.80	&	4.74	&	4.99	&	4.57	\\
HD 165777	&	4.81	&	5.76	&	5.76	&	5.83	&	6.00	&	5.74	\\
HD 173582	&	2.38	&	2.38	&	2.38	&	3.45	&	3.66	&	3.06	\\
HD 173607	&	0.52	&	0.83	&	0.83	&	1.13	&	1.89	&	1.42	\\
HD 173880	&	3.89	&	4.62	&	4.62	&	5.38	&	5.56	&	5.08	\\
HD 174602	&	1.69	&	1.69	&	2.93	&	2.93	&	3.65	&	2.98	\\
HD 177724	&	5.24	&	5.33	&	5.37	&	5.61	&	6.03	&	5.75	\\
HD 184006	&	4.81	&	5.18	&	5.18	&	5.42	&	5.73	&	5.30	\\
HD 192640	&	3.78	&	3.78	&	4.57	&	4.80	&	5.02	&	4.41	\\
HD 199254	&	2.89	&	3.88	&	4.03	&	4.03	&	4.51	&	3.85	\\
HD 204414	&	3.48	&	3.48	&	4.58	&	4.62	&	4.88	&	4.21	\\
HD 205835$^{a}$	&	3.58	&	3.58	&	4.18	&	4.18	&	4.72	&	4.41	\\
HD 210715$^{a}$	&	4.14	&	4.14	&	4.57	&	4.57	&	5.01	&	4.41	\\
HD 213558	&	4.42	&	4.74	&	4.74	&	5.38	&	5.53	&	4.95	\\
\enddata
\tablenotetext{a}{Binary detections. Limits derived after removal of companion from data.}
\label{table5}
\end{deluxetable}

\subsection{Companion Population Analysis} \label{subsec:astar_population}

From the sample in DF 22 and this work, we observed 57 A-type stars in total at CHARA, and we detected seven companions. Three of these targets had interferometric observables that were very noisy due to lacking sufficient calibration sources and were not included in our final demographic analysis, which includes 54 A-type primaries listed in Table \ref{tab:table1}. In our observing approach, we did not observe any dedicated calibrator stars and relied on an expected low companion frequency to then use science targets well-fit by a single star model as calibrators for other science targets observed closely in time.


To characterize the companion population, we perform a Bayesian demographic analysis as in \citet{DeFurio2022_Mstar}, using PyMultiNest \citep{Feroz2009, Buchner2014} to sample the full parameter space, maximize the likelihood function, and derive posteriors for the parameters of our models. See \citet{DeFurio2022_Mstar} for a thorough description of the analysis. We summarize our approach here. In DF22, we characterized the companion population using a frequentist approach that was sensitive over mass ratios = 0.25-1.0 and projected separations = 0.288 - 5.481 au. In this work, we characterize the companion population over mass ratios = 0.1-1.0 and projected separations = 0.01 - 27.54 au having accounted for incompleteness in the Bayesian analysis. This allows us to place constraints over a much larger range of parameter space than previously possible.

First, we assume common functional descriptions of the mass ratio and separation distributions. We model the companion population with two models. The two-parameter model assumes a power-law fit to the mass ratio distribution and a linear flat separation distribution. The four-parameter model assumes a power-law fit to the mass ratio distribution and a log-normal fit to the separation distribution. The power-law mass ratio distribution is defined as:

\begin{equation}
    \Phi(q) \propto
    \begin{cases}
    q^{\gamma} & \text{if }\gamma \geq 0\\
    (1-q)^{-\gamma} & \text{if }  \gamma < 0
    \end{cases}
    \label{qdistribution}
\end{equation}

where $\gamma$ is the power-law index, and the piecewise function ensures symmetry about q=0.5 as defined in \citet{Fontanive2019MNRAS.485.4967F}. The log-normal separation distribution is defined as:

\begin{equation}
\Psi(a) = \frac{1}{\sqrt{2\pi\sigma_{loga}^{2}}} e^{-\frac{(log(a)-log(\Bar{a}))^{2}}{2\sigma_{loga}^{2}}}
    \label{adistribution}
\end{equation}

where $\Bar{a}$ is the mean separation and $\sigma_{loga}$ is the standard deviation of the separation distribution in log-space.

\begin{figure*}
\includegraphics[scale=0.8]{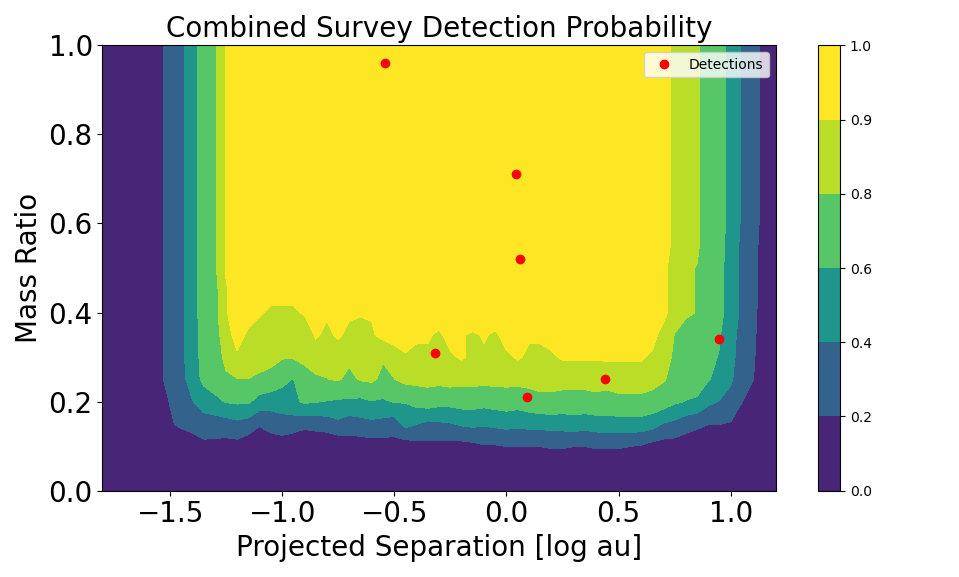}
\caption{Summed detection probabilities for all the 54 sources in our survey.  Red circles show the projected separation and estimated mass ratios for all detected companions. We use this map in Sec. \ref{subsec:astar_population} to model the separation and mass ratio distribution of the companion population to A-type primaries within 80 pc.}
\label{fig:astar_survey_detetionprob}
\end{figure*}

We define the likelihood function from Poisson statistics and the physical parameters of our detected companions \citep{Fontanive2018}. The Poisson likelihood is:

\begin{equation}
\mathcal{L}_{p} = \frac{k^{d} e^{-k}}{d!}
    \label{poissonlike}
\end{equation}

where \textit{k} is the expected number of companion detections given the model and \textit{d} is the number of true observed companions. The expected number of detections is defined as: 

\begin{equation}
k = \sum_{i=1}^{n} p_{i}*CF*\frac{N}{n}
    \label{expected}
\end{equation}

where \textit{CF} is the companion frequency of the sample over mass ratios = 0.1-1 and 0.01-27.54 au, \textit{N} is the total number of sources in our sample (here 54), \textit{n} is the total number of generated companions in the artificial population sampling process, and \textit{p$_{i}$} is the probability that the \textit{i} generated companion will be detected given our survey sensitivity. We constructed the detection probability map by summing over the detection limits of each source in our sample which produces the average sensitivity of the whole survey, see Fig. \ref{fig:astar_survey_detetionprob}. Due to the different distances to our sources, the inner and outer working angle is not the same in terms of physical separation for each source and therefore there is a gradient of sensitivity up to the region of common sensitivity of all sources. We then assign \textit{p$_{i}$} from this map as the probability that the generated companions would be detected given their physical parameters. 

We assume flat priors for -5.0 $<$ $\gamma$ $<$ 5.0 and 0.0 $<$ \textit{CF} $<$ 1.0 and log-flat priors for 0.0 $<$ log(a$_{o}$) $<$ 4.0 and 0.1 $<$ $\sigma_{loga}$ $<$ 5.0 so that each au is equally weighted.  Then, we generate the mass ratio and projected orbital separation distributions of eqs. \ref{qdistribution} and \ref{adistribution} based on each sampled value. From each model, we generate n = 10$^{4}$ companions, with mass ratios = 0.1 - 1 and projected orbital separations = 0.01-27.54 au, and calculate \textit{p$_{i}$} for each companion.  We then sum over all \textit{n} generated companions to evaluate \textit{k}, and compute the Poisson likelihood where \textit{CF} is a free parameter.

We also calculate the likelihood of the model given the physical parameters of our real detections.  First, we generate the model of the population and convolve it with the combined survey detection probability to make a joint probability distribution that gives the expected companion distribution based on the sampled model and the sensitivity of the survey. Finally, we evaluate the joint probability of each true detection, \textit{p$_{j}$}, based on their derived physical parameters.  The total likelihood is calculated as:

\begin{equation}
\mathcal{L} = \mathcal{L}_{p} * \prod_{j=1}^{d} p_{j}
    \label{likelihood}
\end{equation}

\begin{figure*}
\includegraphics[scale=0.65]{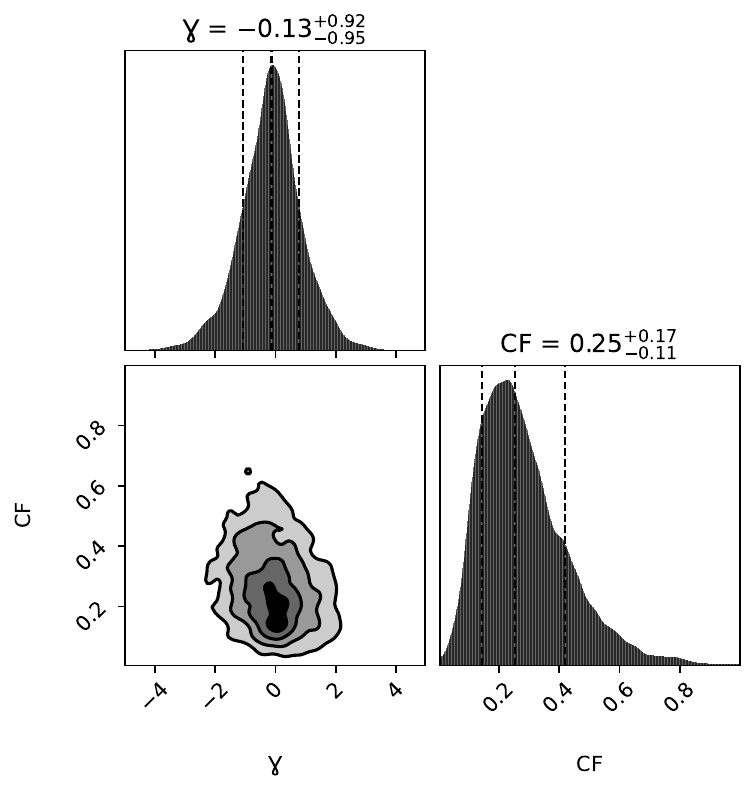}
\includegraphics[scale=0.4]{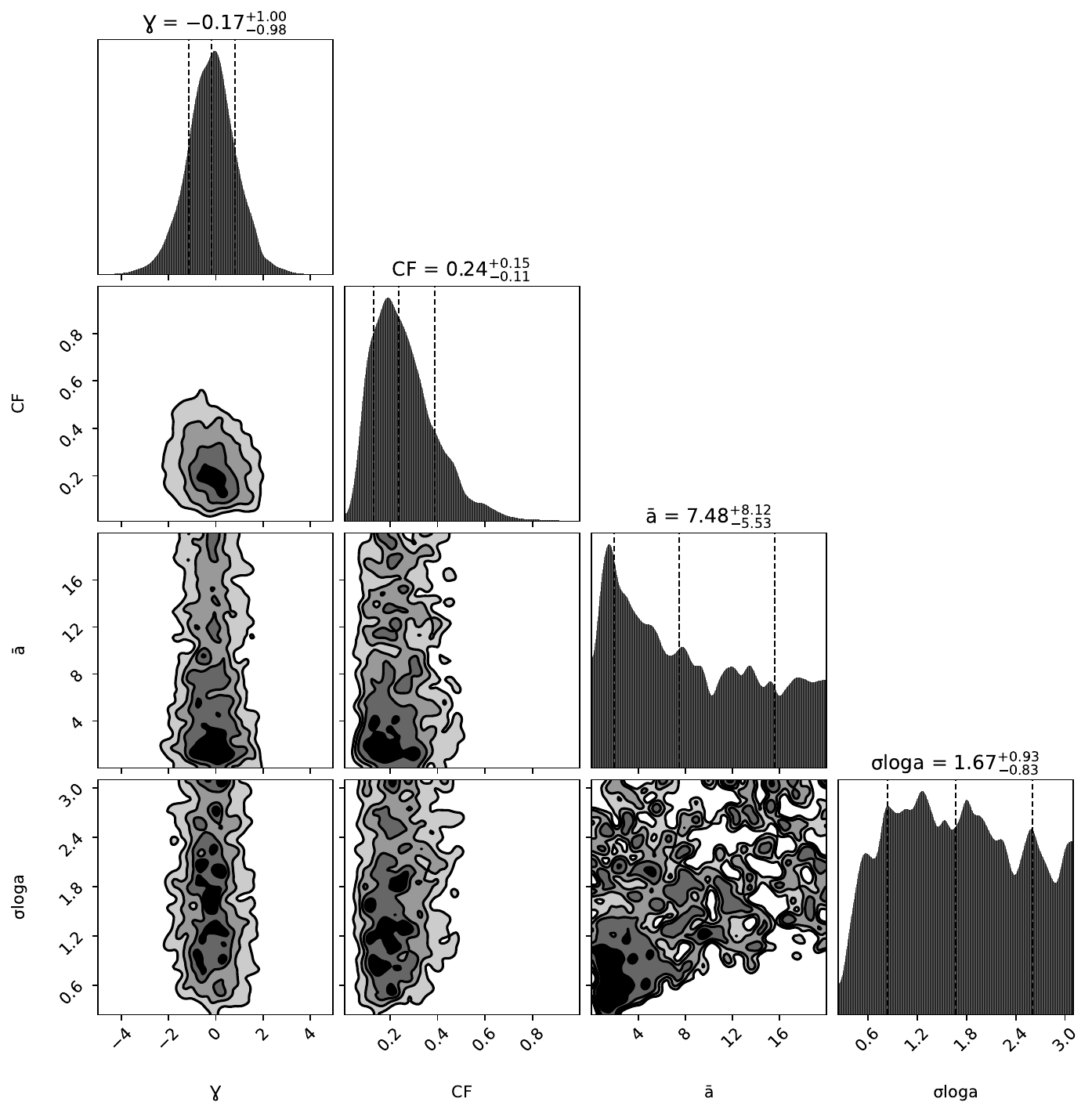}
\caption{Corner plots representing the posterior distributions of the companion population model used in our fit. $\gamma$ is the exponent to the power law model of the mass ratio distribution, CF is the companion frequency over q=0.1-1 and a=0.01-27.54 au, $\Bar{a}$ is the mean separation of the log-normal separation distribution in units of au, and $\sigma_{log(a)}$ is the standard deviation of the log-normal separation distribution. The separation distribution is unconstrained with a log-normal, likely due to the low number of detections in our survey.}
\label{fig:cornerplots_astar}
\end{figure*}

From our sample of 54 A-type primaries, we estimated the following four parameters over mass ratios = 0.1-1.0 and projected separation 0.01-27.54 au with 1$\sigma$ errors (68\% confidence interval): $\gamma$ = -0.17$^{+1.00}_{-0.98}$, \textit{CF} = 0.24$^{+0.15}_{-0.11}$, $\overline{a}$ = 7.48$^{+8.12}_{-5.53}$ au, and $\sigma_{loga}$ = 1.67$^{+0.93}_{-0.83}$. As shown in Fig. \ref{fig:cornerplots_astar}, the parameters of the projected orbital separation distribution are unconstrained, likely due to the low number of detections in our sample. For the two parameter model, we find: $\gamma$ = -0.13$^{+0.92}_{-0.95}$, \textit{CF} = 0.25$^{+0.17}_{-0.11}$. Between these two models, the difference in the Bayesian evidence is 0.2. Therefore, the comparison is inconclusive and we draw no distinction between the two models. This allows us to conclude that the observed companion population is not better modeled by a log-normal distribution fit compared to a flat separation distribution. This is potentially due to the low number of detections or that the underlying separation distribution is wide and that our survey over 0.01-27.54 au (3.44 units in log-separation) does not sample enough of the distribution to constrain the underlying model. We use the two-parameter model for further analysis.

\section{Discussion} \label{sec:astar_discussion}

Our total source list was derived from the \citet{DeRosa2014} VAST sample.  We excluded sources with known Ap or Am spectral types in order to have a sample of chemically typical A-type stars, but did not make sample selections based on previously detected companions. These exclusions may have an impact due to their different spin rate than typical A-type stars, but evidence has been found for an increase in companion frequency for both slow rotators \citep{Smith2024ApJ...975..153S} and fast rotators \citep{Kounkel2023AJ....165..182K}. Therefore, these results should be interpreted for the population of chemically typical A-type stars. Sources were then randomly selected based on the observing nights allocated. Both additional binary detections in this work were previously targeted for companion search through adaptive optics imaging and common proper motion analysis. Only one was found to have a companion over separations sampled, HD 205835, at a separation of 454 au and mass ratio of 0.05. Therefore, this is a triple system with a close binary pair separated by 1.12 au and a widely separated low-mass tertiary, a common configuration for triples. HD 205835 was observed from 32 - 794 au and 3980 - 45000 au. HD 210715 was observed from 32 - 794 au.

\subsection{Comparing the Companion Population to Models} \label{subsec:astar_comparison}

\subsubsection{Mass Ratio Distribution} \label{subsubsec:astar_comparison_massratio}

We derived a best fit power-law index of -0.13$^{+0.92}_{-0.95}$ to the mass ratio distribution for the close companions to A-stars. The \citet{DeRosa2014} multiplicity survey of A-type stars is sensitive to companions beyond $\sim$ 20 au and mass ratios $\geq$ 0.15 for a majority of their sample. For companions between 30-125 au and 125 - 800 au, they describe the mass ratio distribution as a power law with a best-fit index $\gamma$ = -0.5$^{+1.2}_{-1.0}$ and -2.3$^{+1.0}_{-0.9}$ respectively. Our results are consistent with both of these power-law fits given the relatively large errors in their estimate and in our own. A future expansion of this survey would allow for a larger sample of detected companions which would greatly reduce errors on the power-law index to the mass ratio distribution.

\subsubsection{Companion Frequency} \label{subsubsec:astar_comparison_companionfrequency}

\citet{Murphy2018MNRAS.474.4322M} conducted a survey of $\delta$ Scuti variable stars using the phase modulation technique on Kepler light curves, and found a CF = 0.139 $\pm$ 0.021 over $\sim$ 0.6-3.6 au and all mass ratios. Roughly seventy percent of their companions have mass ratios = 0.2-1.0, resulting in a CF = 0.10$\pm$ 0.021 over those mass ratios and separations. In our full sample, we detect four companions out of the 54 sources over the same mass ratios and separations, resulting in a CF = 0.07$^{+0.05}_{-0.02}$ consistent with the \citet{Murphy2018MNRAS.474.4322M} result.

Other surveys have attempted to characterize the companion frequency of A-type stars using LBI with VLTI/Gravity and common proper motion analyses \citep{Waisberg2023MNRAS.521.5232W}. However, they only analyzed stars with high accelerations between epochs of Hipparcos and Gaia which introduces a significant bias in their sample. All high acceleration stars should have a companion over the parameter space sampled and cannot be representative of the A-type population itself. Therefore, we cannot compare our results to their estimated companion frequency, which is heavily biased towards the presence of companions.

Other surveys have characterized the companion population to various types of primary stars. \citet{Raghavan2010} characterized the companion population to Galactic field FGK stars over all separations, \citet{DeRosa2014} to field A-type stars with adaptive optics beyond $\sim$ 20 au, and \citet{Rizzuto2013MNRAS.436.1694R} to B-type stars in Sco-Cen over nearly all separations. These surveys each fit the mass ratio and separation distributions with a power law and a log-normal function, respectively. We use eqs. \ref{qdistribution} and \ref{adistribution} to define the mass ratio and separation distributions and integrate these functions over the sensitivity of our own survey using the stated values of the parameters from past studies to estimate the expected companion frequency for each of these types of primary stars:

\begin{equation}
    CF = C_{n}*\int_{q1}^{q2} \Phi(q)dq \int_{a1}^{a2} \Psi(a)da
    \label{astar_cf}
\end{equation}

\begin{figure}
\includegraphics[scale=0.55]{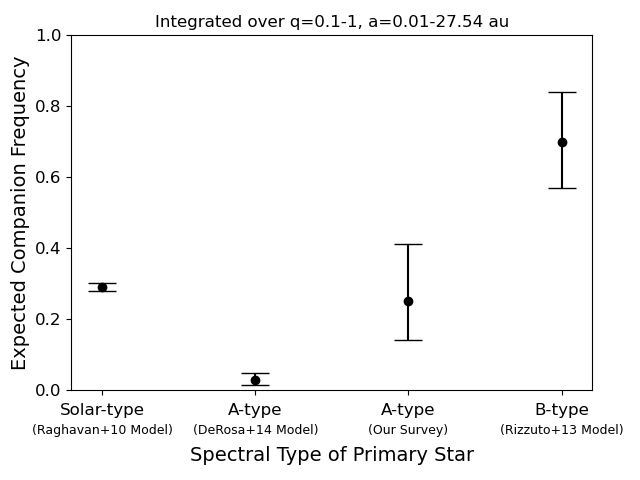}
\caption{Companion frequencies based on spectral type of the primary star from the various models listed in Sec. \ref{subsubsec:astar_comparison_companionfrequency}. All frequencies are calculated over mass ratios = 0.1 - 1.0 and separations (a) = 0.01 - 27.54 au.  The companion frequency for Solar-type primaries is derived from the model of \citet{Raghavan2010}, A-type primaries from the model of \citet{DeRosa2014} and this survey, and B-type primaries from the model of \citet{Rizzuto2013MNRAS.436.1694R}.}
\label{fig:astar_cfs}
\end{figure}

We first integrate these functions over the sensitivity of their own survey sampling the values of CF, $\gamma$, $\sigma_{loga}$, and $\overline{a}$ 10$^{5}$ times, see Table \ref{table6}. Each evaluation of eq. \ref{astar_cf} allows us to solve for C$_{n}$ which we then input back into the same equation, but this time integrating over the sensitivity of our survey in this work, q=0.1-1.0 and a=0.01-27.54 au, to arrive at a value for CF, see Fig. \ref{fig:astar_cfs}. We lastly evaluate the posterior probability that the expected CF of each of these surveys over our sensitivity is the same as that of our survey. We integrate the posterior distribution of the companion frequency from our two parameter model, see Fig. \ref{fig:cornerplots_astar}, from 0 to the CF of each sampling stated above for the A-type model and from the CF of each sampling to infinity for the FGK- and B-type models. This ensures that the correct tail of the posterior distribution is evaluated to determine the posterior probability that the model can describe our observations of A-stars. In Table \ref{table6}, we present the estimated posterior probabilities for the companion population models to solar-type, A-type, and B-type primaries which are 0.53, 0.005, and 0.04 respectively. From these results, we find that: a) there is no evidence for a difference between the multiple population of our A-type sample and that of solar-type stars as found in DF22, b) the companion frequency for A-type stars at closer separations is higher than if extrapolating the \citet{DeRosa2014} log-normal separation distribution which should apply for separations greater than roughly 100 au where they are mostly complete, and c) the companion frequency to B-type primaries is potentially larger than A-type primaries over mass ratios and separations sampled which was not found in DF22 due to the low sample size and much more stringently probed parameter space utilizing the frequentist approach.

\begin{deluxetable*}{ccccccccc}
\tablenum{6}
\tablecolumns{9} 
\tablecaption{Companion population parameters and sensitivities of each tested model with the resulting statistics.}
\tablehead{\colhead{Primary} &\colhead{$\gamma$} &\colhead{log(a$_{o}$)}  &\colhead{$\sigma_{loga}$} & \colhead{CF}& \colhead{log(a)}& \colhead{q} &\colhead{Expected CF} &  Posterior \\ \colhead{Spectral Type} & \colhead{}&\colhead{}&\colhead{}&\colhead{}&\colhead{sensitivity}&\colhead{sensitivity} &\colhead{} &\colhead{Probability} }
\startdata
FGK-type$^{1}$ & 0 & 1.7 & 1.68 & 0.61 $\pm$ 0.02 & -2 $\leq$ log(a) $\leq$ 4 & q $\geq$ 0.1 & 0.29 $\pm$ 0.01 & 0.53 \\
A-type$^{2}$ & -0.5$^{+1.2}_{-1.0}$ & 2.59 $\pm$ 0.13 & 0.79 $\pm$ 0.12 & 0.219 $\pm$ 0.026 & 1.5 $\leq$ log(a) $\leq$ 2.9 & q $\geq$ 0.1 & 0.028$^{+0.018}_{-0.014}$ & 0.005\\
B-type$^{3}$ & -0.46 $\pm$ 0.14 & 1.05 $\pm$ 0.2 & 1.35 $\pm$ 0.2 & 1.35 $\pm$ 0.20 &  -2 $\leq$ log(a) $\leq$ 4 & q $\geq$ 0.1 & 0.70$^{+0.14}_{-0.13}$ & 0.04 \\
\enddata
\tablenotetext{1}{\citet{Raghavan2010}}
\tablenotetext{2}{\citet{DeRosa2014}}
\tablenotetext{3}{\citet{Rizzuto2013MNRAS.436.1694R}}
\label{table6}
\end{deluxetable*}

\subsection{Implications} \label{subsec:implications}

Our analysis of 54 chemically typical A-type stars within 80 pc observed with long-baseline interferometry reveals a companion frequency of 0.25$^{+0.17}_{-0.11}$ over mass ratios 0.1-1.0 and projected separations 0.01-27.54 au.  As in DF22, we find that extrapolating the companion population model of \citet{DeRosa2014} cannot reproduce our observed results, see Fig. \ref{fig:astar_cfs}. The best-fit power law index to the mass ratio distribution shows hints that it is consistent with the close companion population from \citet{DeRosa2014}, but we cannot rule out the power law index of the wide companion population which is within 2$\sigma$. We do not find any difference between the companion frequency of solar-type stars in the field \citep{Raghavan2010} and our sample of A-type stars, over mass ratios = 0.1-1.0 and projected separations = 0.01-27.54 au. However, we calculate a probability of 0.04 that the companion population of B-type stars can describe the observations of the companion population of our A-star sample due to the abundance of close-in companions to B-type primaries.
See Table \ref{table6} and Fig. \ref{fig:astar_cfs} for a comparison to other surveys over the sensitivity of our survey.

It is well established in the literature that more massive stars have a higher companion frequency, and that there is a break in the separation distributions: low-mass and solar-type stars follow a log-normal with increasingly fewer companions at closer separations, whereas high companion frequencies of high mass stars continue towards closer separations \citep{Moe2017}. However, the trend in the companion frequency at close separations is unclear for A-type primaries. Our results suggest that intermediate-mass A-type binaries are more like solar-type stars and have similar formation pathways. Primaries with $M \gtrsim 3M_\odot$ (O and B stars) follow similar distributions as more massive stars. This break could be consistent with an enhanced role for disk fragmentation only for high-mass primaries, which is consistent with theoretical predictions \citep{Kratter2006MNRAS.373.1563K}. Numerical simulations are able to reproduce even close-in low-mass binaries through turbulent fragmentation coupled with migration and dynamical interactions \citep{Bate2018MNRAS.475.5618B}. Similarly, large star formation simulations like STARFORGE, which better sample the IMF \citep{Grudic2022MNRAS.512..216G, Guszejnov2023MNRAS.518.4693G}, are able to recover the multiplicity properties of low mass stars reasonably well despite not having resolved protostellar disks. However, they somewhat underproduce companions at higher masses, perhaps due to the lack of resolved disk fragmentation. Note that there remain uncertainties in how to compare the simulated and observed companion fractions due to incompleteness at the lowest masses, and thus mass ratios, as well as closest separations.
Ultimately, with a larger sample of A-type stars, we will be able to explore differences in the companion population as a function of primary mass and differentiate between potential formation mechanisms through the derived mass ratio distribution.

\section{Conclusion} \label{sec:conclusion}
We have conducted a multiplicity survey of 54 nearby, chemically typical A-type stars within 80 pc using long baseline interferometry at the CHARA Array with the MIRC-X and MYSTIC instruments. To summarize the results of our survey:

1) We detected seven companions in total, two from the newest data, with projected separations 6-158 mas (0.287-8.747 au) and mass ratios = 0.21-0.96. For our sample of A-type stars with masses of 1.44-2.93 M$_{\odot}$, we observed a completeness-corrected companion frequency of 0.25$^{+0.17}_{-0.11}$ over the sensitivity of our survey, mass ratios 0.1-1.0 and projected separations 0.01-27.54 au.

2) Our results are consistent with what we had previously found in DF22, except now characterized over a much larger range in parameter space due to the inclusion of a Bayesian demographic analysis.

3) Our estimate of the companion frequency is consistent with that of FGK-type stars over the sensitivity sampled, but we find a probability of 0.04 that the B-type companion population model can replicate our observed companion frequency to A-type primaries over the sensitivity of our survey. This may be indicative of an increased occurrence of companions to B-type stars ($\gtrsim$ 2.5 M$_{\odot}$), potentially due to a higher incidence of disk fragmentation.

\section{acknowledgments}
This work is based upon observations obtained with the Georgia State University Center for High Angular Resolution Astronomy Array at Mount Wilson Observatory.  The CHARA Array is supported by the National Science Foundation under Grant No. AST-2034336 and AST-2407956. Institutional support has been provided from the GSU College of Arts and Sciences and the GSU Office of the Vice President for Research and Economic Development. Time at the CHARA Array was granted through the NOIRLab community access program (NOIRLab PropID: 2020B-0290; PI: M. De Furio). MIRC-X received funding from the European Research Council (ERC) under the European Union's Horizon 2020 research and innovation programme (Grant No. 639889). S.K. acknowledges funding for MIRC-X received funding from the European Research Council (ERC) under the European Union's Horizon 2020 research and innovation programme (Starting Grant No. 639889 and Consolidated Grant No. 101003096). JDM acknowledges funding for the development of MIRC-X (NASA-XRP NNX16AD43G, NSF-AST 2009489) and MYSTIC (NSF-ATI 1506540, NSF-AST 1909165). This research has made use of the Jean-Marie Mariotti Center \texttt{Aspro} service, available at http://www.jmmc.fr/aspro. This research has made use of the Jean-Marie Mariotti Center \texttt{SearchCal} service, which involves the JSDC and JMDC catalogues, available at https://www.jmmc.fr/searchcal. This work has made use of data from the European Space Agency (ESA) mission {\it Gaia} (\url{https://www.cosmos.esa.int/gaia}), processed by the {\it Gaia} Data Processing and Analysis Consortium (DPAC, \url{https://www.cosmos.esa.int/web/gaia/dpac/consortium}). Funding for the DPAC has been provided by national institutions, in particular the institutions participating in the {\it Gaia} Multilateral Agreement. M.D.F. is supported by an NSF Astronomy and Astrophysics Postdoctoral Fellowship under award AST-2303911. M.D.F. acknowledges S.F. KMK acknowledges support from the NSF AAG program under award 2407522.

\bibliography{bibliography}{}
\bibliographystyle{aasjournal}

\end{document}